\documentclass[twocolumn,showpacs,amsmath, amssymb, prd]{revtex4-1}
\usepackage[pdftex]{graphicx}
\usepackage{dcolumn}
\usepackage{bm}

\begin{document}

\title{Pulsations in short gamma ray bursts from black hole-neutron star mergers}
\author{Nicholas Stone, Abraham Loeb, and Edo Berger}
\email{nstone@cfa.harvard.edu}
\affiliation{Astronomy Department, Harvard University, 60 Garden Street,
Cambridge, Massachusetts 02138, USA}

\begin{abstract}

The precise origin of short gamma ray bursts (SGRBs) remains an important open question in relativistic astrophysics.  Increasingly, observational evidence suggests the merger of a binary compact object system as the source for most SGRBs, but it is currently unclear how to distinguish observationally between a binary neutron star progenitor and a black hole-neutron star progenitor.  We suggest the quasiperiodic signal of jet precession as an observational signature of SGRBs originating in mixed binary systems, and quantify both the fraction of mixed binaries capable of producing SGRBs and the distributions of precession amplitudes and periods.  The difficulty inherent in disrupting a neutron star outside the horizon of a stellar mass black hole (BH) biases the jet precession signal towards low amplitude and high frequency.  Precession periods of $\sim 0.01-0.1~{\rm s}$ and disk-BH spin misalignments $\sim 10^{\circ}$ are generally expected, although sufficiently high viscosity may prevent the accumulation of multiple precession periods during the SGRB.  The precessing jet will naturally cover a larger solid angle in the sky than would standard SGRB jets, enhancing observability for both prompt emission and optical afterglows.

\end{abstract}


\date{\today}
\maketitle

\section{Introduction}
The origin of the short-hard gamma-ray bursts (SGRBs; durations $T_{90}\lesssim 2$ s; \citep{kmf+93}) was largely a matter of
speculation until the recent discovery of their afterglows and host galaxies (e.g., Refs. \citep{bpc+05,ffp+05,gso+05,bpp+06}).  These observations have demonstrated that SGRBs are cosmological in origin ($z\gtrsim 0.1$; \citep{bfp+07}); have a beaming-corrected energy scale of $\sim 10^{49}-10^{50}$ erg \citep{bgc+06,sbk+06,Fong12}; lack associated supernovae (e.g., Refs. \citep{hwf+05,sbk+06}); occur in a mix of star-forming and elliptical galaxies \citep{ber09}; have a broad spatial distribution around their hosts \citep{FBF}, with some events offset by tens of kpc \citep{ber10}; and have low-density parsec-scale environments \citep{sbk+06,FBF}.  The confluence of these characteristics provides support to the popular model of compact object (CO) mergers \citep{Paczynski91}.

In this context, the key open question that motivates our paper is the following: if SGRBs originate in CO mergers, what types of compact objects are merging?  Specifically, a neutron star (NS) can theoretically be tidally disrupted by, and produce an accretion disk around, either a more compact NS, or a sufficiently small stellar mass black hole (BH).  Unfortunately, it is not clear how to distinguish between mergers of neutron star binaries (NS-NS) and mixed binaries (BH-NS).  The advent of gravitational wave (GW) astronomy will facilitate this task, but distinguishing the waveforms accompanying NS-NS mergers from those in BH-NS mergers is a nontrivial task \citep{HBF13}, and furthermore the Advanced LIGO era is still half a decade away.  It is also possible that GW signals seen by Advanced LIGO will lack accompanying electromagnetic counterparts if the SGRB beaming angle is too small, or if the intrinsic event rate is too low.  In this paper we suggest a clear observational tool for distinguishing between NS-NS and BH-NS progenitors of SGRBs using electromagnetic data only: the precession of the disks and jets associated with these events.

Jet precession has previously been discussed as a phenomenon relevant for SGRBs, originally with regard to now-disfavored SGRB models \citep{RFT, BIF} but later in the context of CO mergers.  Early works in the CO merger paradigm considered disks fed by stable mass transfer from a NS onto a BH so that the precession was forced by tidal torques \citep{PZLL}, but subsequent models considered the more realistic neutrino-dominated accretion flows (NDAFs) formed after tidal disruption of a NS by a stellar mass BH \citep{RRS}.  This last model is similar to the one presented in this paper, in that it considers a thick disk precessing as a solid body rotator due to general relativistic Lense-Thirring torques.  It has been applied to predict different observational signatures, such as the light curves of precessing jets \citep{RRS, LWG} or even LIGO-band GW signals emitted by forced precession of a large amount of disk mass \citep{RRC, SLGL}.  Others have considered an inclined disk whose inner regions include a Bardeen-Petterson warp; in their model the inner region of a NDAF precesses along with the BH spin about the total angular momentum vector \citep{Liu10}.  

The above papers generally treat disk precession in an analytic or semianalytic way, due to the high computational expense of even Newtonian simulations.  However, the late inspiral, plunge, and merger of a BH-NS binary has been repeatedly simulated in full numerical relativity (see Ref. \citep{ST11} for a review), providing a detailed picture of the initial conditions dictating subsequent disk evolution.  Shortly after the first numerical relativity simulations of a BH-BH merger \citep{P05}, fully relativistic BH-NS mergers were simulated by multiple groups \citep{TBF06, G06, SU06, SU07}.  The results of these early simulations were later refined, and subsequent work considered the effects of varied mass ratios \citep{EFL08, SKY09, KOS11}, BH spin \citep{ELS09, FDK, KOS11}, NS compactness and equation of state \citep{KST10, PRO11, FDD12}, spin-orbit misalignment \citep{FDKT, FDD12}, orbital eccentricity \citep{EPS12b}, and magnetic fields \citep{ELP12, EPS12}.  Although disk precession has been seen in misaligned BH-NS merger simulations \citep{FDKT}, the semianalytic approach of this paper has two main advantages over numerical simulation: we both consider the evolution of the remnant accretion disk over long time scales, and finely sample the broad parameter space of these events.

This paper follows Ref. \citep{RRS} in focusing on thick disks precessing as solid body rotators, which are well motivated for the supercritical accretion flows and misaligned angular momentum vectors characteristic of compact object mergers (Sec. II).  Our work differs from past efforts, however, in our consideration of the viscous spreading of the disk, as well as our adoption of simplifying assumptions tailored to match results from numerical relativity (NR) simulations of mixed binary mergers (Sec. III).  We quantify for the first time the distributions of precession periods and angles given physically motivated assumptions about progenitor spins and masses (Sec. IV), using simple analytic formulas when appropriate and more complex empirical fits to NR simulations when necessary.  In the process we estimate the fraction of BH-NS mergers that can actually produce accretion disks.  We conclude by considering the observable consequences of jet precession in the context of SGRBs (Sec. V).  Unlike in previous work, we discuss both the case where the jet is tied to the BH spin vector, and the case where it aligns with the disk angular momentum vector.

\section{Disk Precession}

Nonaxisymmetric torques will, initially, induce small warps in accretion disks due to differential precession between adjacent mass annuli.  The evolution of these warps depends on how they are able to propagate through the disk.  When the disk is sufficiently viscous, warps propagate diffusively, allowing differential precession to produce substantial shear viscosity, and dissipating large amounts of orbital energy.  In the context of tilted accretion disks experiencing Lense-Thirring torques from BH spin, diffusive propagation of warps will align the inner regions of the disk with the black hole equatorial plane; this is known as the Bardeen-Petterson effect \citep{BP, PP83, Ogilvie99}.

In the opposite regime, a thick disk with a short sound-crossing time scale will propagate warps in a wavelike manner, redistributing torques throughout the disk and inducing near-rigid body precession \citep{PL95, PT95}.  In particular, rigid body precession is possible if $H/r>\alpha$, where $H$ is the disk height and $\alpha$ the dimensionless viscosity parameter at a radius $r$.  This is the regime most relevant for compact object mergers, and is therefore what we will consider for the remainder of this paper.  Approximately rigid body precession has been seen in hydrodynamical simulations of protoplanetary disks being torqued by a binary companion \citep{LNPT}, in GRMHD simulations of tilted accretion disks around spinning black holes \citep{FBAS, FB08}, and, notably, in NR simulations of BH-NS mergers \citep{FDKT}.

In the Newtonian limit, a solid body rotator will precess with a period $T_{\rm prec}=2\pi \sin{\psi_{\rm d}}(J/\mathcal{N})$, where $\psi_{\rm d}$ is the misalignment angle between the accretion disk and the BH equatorial plane, $J$ is the total angular momentum of the disk, and $\mathcal{N}$ is the Lense-Thirring torque integrated over the entire disk.  Specifically, if the disk possesses a surface density profile $\Sigma( r)$ that is nonzero between an inner radius $R_{\rm i}$ and an outer radius $R_{\rm o}$, and the disk elements possess orbital frequency $\Omega ( r)$, then 
\begin{equation}
J=2\pi \int_{R_{\rm i}}^{R_{\rm o}} \Sigma( r) \Omega( r) r^3 {\rm d}r,
\end{equation}
and
\begin{equation}
\mathcal{N}=4\pi \frac{G^2M_{\rm BH}^2a_{\rm BH}}{c^3}\sin\psi_{\rm d} \int_{R_{\rm i}}^{R_{\rm o}} \frac{1}{r^3}\Sigma( r) \Omega( r) r^3 {\rm d}r,
\end{equation}
where the BH's mass and dimensionless spin are $M_{\rm BH}$ and $a_{\rm BH}$, respectively.  Note that in this section $M_{\rm BH}$ and $a_{\rm BH}$ refer to postmerger BH quantities; from Sec. III onward we will distinguish these from the mass and spin of the premerger BH.  Throughout this paper $G$ is the gravitational constant and $c$ is the speed of light.  If we use the Keplerian orbital frequency $\Omega_{\rm k}=\sqrt{GM_{\rm BH}/r^3}$, then for a density profile of the form $\Sigma( r) = \Sigma_0 (r/r_0)^{-\zeta}$, the precession time scale is
\begin{equation}
T_{\rm prec}=\frac{\pi R_{\rm g}(1+2\zeta)}{c(5-2\zeta)}\frac{r_{\rm o}^{5/2-\zeta}r_{\rm i}^{1/2+\zeta}(1-(r_{\rm i}/r_{\rm o})^{5/2-\zeta})}{a_{\rm BH}(1-(r_{\rm i}/r_{\rm o})^{1/2+\zeta})}. \label{Tprec}
\end{equation}
Here we have normalized $r_{\rm o}=R_{\rm o}/R_{\rm g}$ and $r_{\rm i}=R_{\rm i}/R_{\rm g}$ by the gravitational radius $R_{\rm g}=GM_{\rm BH}/c^2$.  Other effects that would influence Eq. \eqref{Tprec} include nutation and relativistic corrections to the orbital frequency, which we neglect in this analysis (but see Refs. \citep{PZLL, RRS}). 

We stress that $T_{\rm prec}$ is an instantaneous precession period, and that for the nonequilibrium disks expected in CO mergers important quantities such as $\zeta$ and $r_{\rm o}$ will be time dependent (although the cancellation of $\Sigma_0$ in $J/\mathcal{N}$ means that the secular decrease in disk mass will {\it not} affect $T_{\rm prec}$).  The time dependence of these variables will cause any signal to be quasiperiodic rather than periodic.  Because the dominant feature of the disk's dynamical evolution will be viscous outward spreading \citep{MPQ08, MPQ09}, we expect $T_{\rm prec}$ to increase with time.

Although a power law definition of $\Sigma$ is appealing for its simplicity, both analytical models \citep{MPQ09} and NR simulations \citep{FDKT} indicate that the true structure of these disks is more complex.  To better account for realistic disk structure, and also to quantify the time evolution of the disk as it spreads outward, we adopt the SGRB disk model of Ref. \citep{MPQ08}, which derives exact $\Sigma$ solutions for a viscously spreading ring of matter, and then couples these solutions to more detailed models of disk energetics and composition.  As matter from the disk accretes onto the BH, the intially advective NDAF will become optically thin to neutrinos and geometrically thinner.  At later times ($t\gtrsim 0.1$ sec) the disk will become a geometrically thick, radiatively inefficient accretion flow.  In principle, the intermediate neutrino-cooled period could prevent later disk precession by aligning the disk into the BH midplane through the creation of a Bardeen-Petterson warp.  In practice, it seems that even the neutrino-cooled phase of accretion still possesses $H/r>\alpha$, and is therefore unlikely to align (Ref. \citep{MPQ09}, Fig. 2).

We are primarily interested in the radiatively inefficient accretion flow stage, both because it has the longest duration, and because for low disk masses ($M_{\rm d}\lesssim 0.1M_{\rm NS}$, which is the case for most BH-NS mergers - see Sec. III), it is the only phase of accretion.  For this stage of disk evolution, we can write the surface density as
\begin{align}
\Sigma (r, t)=&\frac{M_{\rm dis}(1-n/2)}{\pi R_{\rm dis}^2x^{n+1/4}\tau}\exp\left(\frac{-(1+x^{2-n})}{\tau} \right)\label{MetzgerDisk} \\
 &\times I_{1/|4-2n|}\left( \frac{2x^{1-n/2}}{\tau} \right) \notag.
\end{align}
Here $M_{\rm dis}$ is the initial disk mass, $R_{\rm dis}$ is the initial radius of the spreading mass ring (i.e. the radius where the NS is disrupted), $I_m$ is a modified Bessel function of order $m$, $x=r/R_{\rm dis}$, $\tau=t(12\nu_0 (1-n/2)^2/R_{\rm dis}^2)$, and we have assumed viscosity of the form $\nu=\nu_0 x^n$.  We calibrate $\nu_0$ with the initial relation $t_{\rm visc, 0}=R_{\rm dis}^2/\nu$ and the equation
\begin{align}
t_{\rm visc, 0}\approx 0.11 \alpha_{-1}^{-1} M_8^{-1/2} R_{\rm dis, 5}^{3/2} \times \left(\frac{H_0}{0.3R_{\rm dis}} \right)^{-2}~{\rm s}, 
\end{align}
where $\alpha$ is the dimensionless Shakura-Sunyaev viscosity coefficient and $H_0$ is the characteristic disk height.  We use the normalizations $\alpha_{-1}=\alpha/0.1$, $M_8=M_{\rm BH}/8M_{\odot}$, and $R_{\rm dis, 5}=R_{\rm dis}/10^5~{\rm m}$.  The value of $\alpha$ is set by the magneto-rotational instability (MRI), and has been estimated to span a wide range of values, from $\sim 0.01$ in local, shearing box simulations \citep{DSP10} to $\sim 1$ in global GRMHD simulations \citep{McKinneyNarayan07}.  However, large $\alpha$ values seen in global simulations are confined to small radii, and in these simulations $\alpha \sim 0.1$ at $r\gtrsim 10r_{\rm g}$.  The importance of $\alpha$ for our results lies primarily in how viscosity controls the outward spreading of the disk, so we follow Ref. \citep{ChenBeloborodov07} and consider large radii $\alpha$ values of $0.01$, $0.03$, and $0.1$.  

$H_0$ will vary both in radius and in time; generally, $H_0$ grows as one moves further out in the disk \citep{ChenBeloborodov07}, and also as the outer edge of the disk viscously spreads, putting a larger fraction of the disk into a purely advective regime with large height \citep{MPQ08}.  Our results are fairly sensitive to both $\alpha$ and $H_0$, but because the former spans a wider range we vary $\alpha$ and fix $H_0=0.3R$.  We arrive at this value by considering the size of the disk at a time $t_{\rm 1/2}$, a characteristic, ``halfway,'' precession time scale.  Specifically, $t_{1/2}=((t_{0}^{-1/3}+1)/2)^{-3}$ is the time at which half the SGRB's precession cycles will have occurred if it lasts from $t=t_{0}$ to $t=1~{\rm s}$ and viscous spreading of the disk causes $T_{\rm prec}\propto t^{4/3}$.  Typically, $t_{\rm 1/2}\sim 100$ msec, which corresponds to a disk outer edge at $r_{\rm o}\sim 50$; at these distances and times, more detailed modeling of disk structure \citep{ChenBeloborodov07} indicates $H/R\approx 0.3$.

\begin{figure}[!t]
\includegraphics[width=85mm]{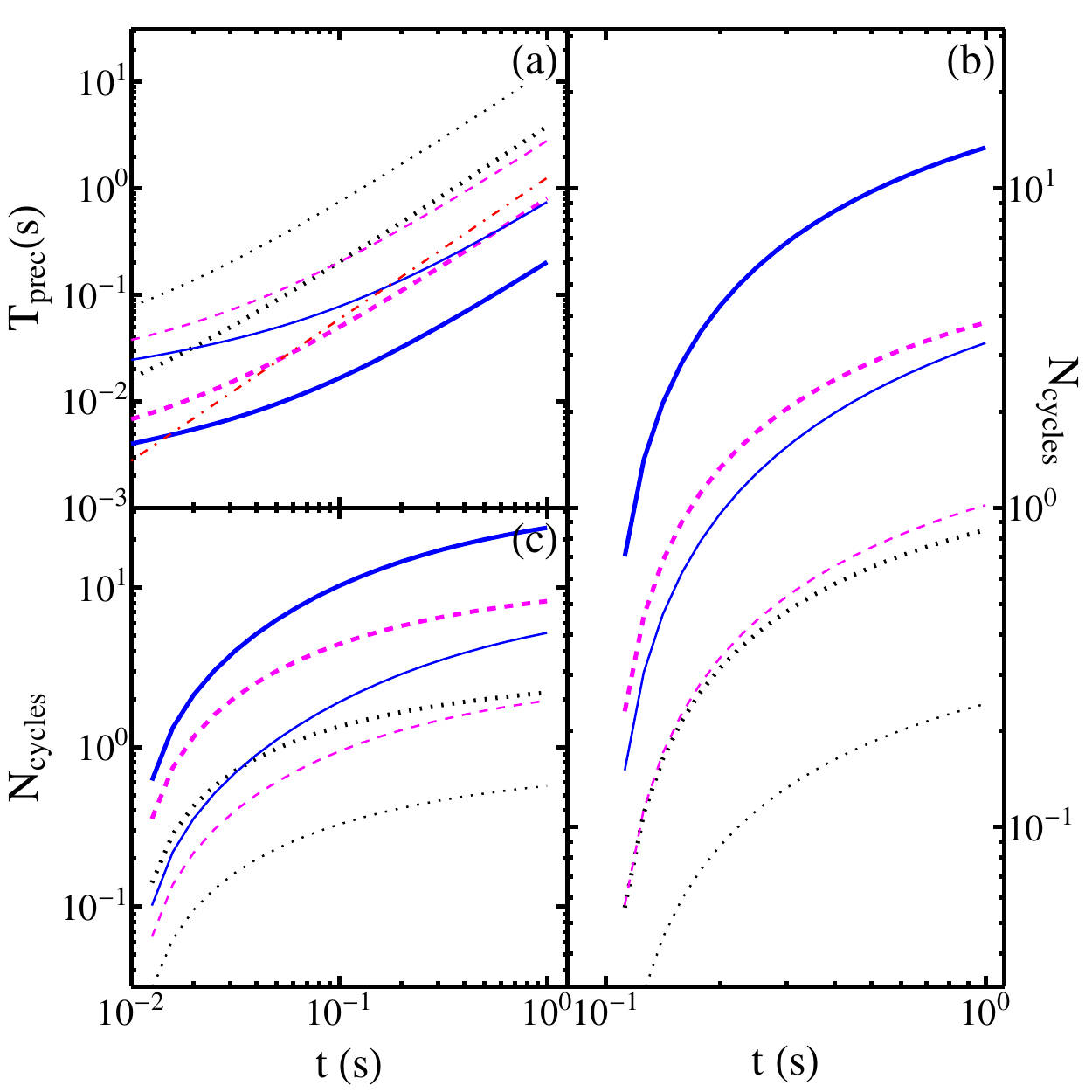}
\caption{(a) Time evolution of $T_{\rm prec}$ assuming a viscously spreading disk structure given by Eq. \eqref{MetzgerDisk}.  Black dotted curves represent $\alpha=0.1$, dashed magenta curves $\alpha=0.03$, and solid blue curves $\alpha=0.01$.  Thick curves are for nearly equatorial disruptions with $a_{\rm BH}=0.9$, while thin curves are for $a_{\rm BH}=0.9$ and initial spin-orbit misalignment of $70^{\circ}$, or equivalently a nearly aligned disruption with $a\approx 0.5$.  The dash-dotted red line is $\propto t^{4/3}$, the rough time evolution of $T_{\rm prec}$.  (b) and (c) show $N_{\rm cycles}$, the accumulated number of cycles for $0.1~{\rm s}<t<1~{\rm s}$ and $0.01~{\rm s}<t<1~{\rm s}$, respectively.}
\label{TprecEv}
\end{figure}

Using Eq. \eqref{MetzgerDisk}, we plot the time evolution of $T_{\rm prec}$ in Fig. \ref{TprecEv}, and find that it increases in rough agreement with analytic expectations: at late times, Eq. \eqref{MetzgerDisk} approaches a power law with $\zeta=1/2$, and the outer edge of the disk expands with time: $r_{\rm o} \propto t^{2/3}$.  The disk mass declines slowly, with $M_{\rm d} \propto t^{-1/3}$, but the mass accretion rate declines faster, with $\dot{M}_{\rm d} \propto t^{-4/3}$.  More specifically, the late-time self-similar solutions are:
\begin{align}
M_{\rm d} &= 0.021 \alpha_{-1}^{-1/3}M_8^{-1/6}R_{\rm dis, 5}^{1/2}M_{\rm dis, -1}t^{-1/3}~M_{\odot} \\
\dot{M}_{\rm d} &= 0.007 \alpha_{-1}^{-1/3} M_8^{-1/6}R_{\rm dis, 5}^{1/2}M_{\rm dis, -1} t^{-4/3}~M_{\odot}{\rm s}^{-1} \label{MDotMetz} \\
R_{\rm o} &= 2.3\times10^{6} \alpha_{-1}^{2/3}M_8^{1/3}t^{2/3}~{\rm m}
\end{align}
Here $M_{\rm dis, -1}=M_{\rm dis}/0.1M_{\odot}$ and $t$ is in units of seconds.  Assuming that $r_{\rm i}$ remains fixed (and ignoring lower-order contributions from $r_{\rm o}$), Eq. \eqref{Tprec} then implies $T_{\rm prec} \propto t^{4/3}$.  In Fig. \ref{TprecEv} we also plot $N_{\rm cycles}$, the total number of precession cycles undergone during the GRB.  For $\alpha \gtrsim 0.1$, a viscously spreading SGRB disk will generally experience $N_{\rm cycles} \lesssim 1$.

In the above discussion we have assumed that the angular momentum lost by inspiralling disk matter is redistributed outward solely by internal viscous torques.  An alternate possibility is specific angular momentum loss through a disk wind, which if magnetized can further torque the disk \citep{MPQ08}.  Adopting the simple ``ADIOS'' model for disk wind losses \citep{BB99}, i.e. $\dot{M}_{\rm d}(r) \propto r^p$ and $\dot{J} = -C\dot{M}_{\rm out}\sqrt{GMR_{\rm o}}$, we can use the similarity solutions of Ref. \citep{MPQ08} for the evolution of an advective disk.  In particular, if the wind is unmagnetized and only specific angular momentum is lost, then $C=2p/(2p+1)$ and $r_{\rm o} \propto t^{2/3}$, just as in the wind free case.  The disk mass and accretion rate will decline more rapidly than in the wind free case, with $M_{\rm d} \propto t^{-1}$ and $\dot{M}_{\rm d} \propto t^{-8/3}$.  

On the other hand, if the outflow exerts a significant magnetic torque on the disk ($C\approx 1$), then $r_{\rm o} \propto t^{1/(p+3/2)}$ and $T_{\rm prec} \propto t^{2/(p+3/2)}$.  Both analytic \citep{Begelman12} and numerical works \citep{HB02} have suggested that a value of $p=1$ is roughly appropriate for radiatively inefficient flows.  This yields $r_{\rm o}\propto t^{2/5}$, $T_{\rm prec} \propto t^{4/5}$, and a rapidly declining disk mass, with $M_{\rm d} \propto \exp(-1.15(t/t_{\rm visc, 0})^{2/5})$ .  If this is the case, $N_{\rm cycles}\gtrsim 10$ for all realistic $\alpha$ values, but the rapid loss of disk mass may make precession more difficult to observe (as discussed in Sec. V).

For the remainder of this paper, however, we conservatively calculate fiducial precession time scales using Eq. \eqref{MetzgerDisk} and $n=1/2$.  This may underestimate $N_{\rm cycles}$, but due to the number of additional free parameters, incorporating a detailed outflow model is beyond the scope of this paper.  Our results are generally insensitive to $R_{\rm dis}$ and $M_{\rm dis}$ (although we will later use the initial disk mass $M_{\rm dis}$ as a criterion for whether or not a SGRB can form - see Sec. III).  Our results are sensitive to $\alpha$, with large $\alpha$ values increasing the precession period and decreasing the number of precession cycles that can fit in the duration of an SGRB.  In all cases we self-consistently calculate the inner disk edge $r_{\rm i}$ using the formalism in Ref. \citep{Gabe12} for finding the innermost stable spherical orbit (ISSO), the tilted analogue to the innermost stable circular orbit (ISCO).  Details of the ISSO calculation are in Appendix A. 

The precession of the SGRB disk in isolation is unlikely to be observable and is mainly interesting as a source of jet precession.  The observational signatures of jet precession will hinge on two uncertain astrophysical questions: the opening angles of SGRB jets and the alignment direction of a jet in a tilted accretion flow.  The first of these questions has recently become amenable to observational constraint; observations of jet breaks in SGRBs suggest opening angles of $\sim 10^{\circ}$ \citep{sbk+06}.  Observational evidence for the second question is limited, and ambiguous.  Observations of a relativistic outflow following tidal disruption of a star by a supermassive BH (Swift J164449.3+573451) suggested that in that case, the jet aligned with the BH spin axis \citep{SL}; on the other hand, observations of the microquasar LSI+61303 have been interpreted as evidence of a precessing jet, aligned with the angular momentum axis of a precessing disk \citep{MRZ}.  There may not be a universal answer to this question, as different hypothetical jet launching mechanisms might each tie the jet axis to a different preferred direction.  

However, for the two leading jet launching mechanism candidates in SGRBs - $\nu \bar{\nu}$ pair annihilation \citep{MeszarosRees92, RuffertJanka99}, and the Blandford-Znajek (BZ) mechanism \citep{BlandfordZnajek77, Lee+00} - there are theoretical reasons to believe that the jet will align with the disk angular momentum vector.  The $\nu \bar{\nu}$ annihilation scenario is independent of BH spin and depends only on disk properties.  Alignment of a BZ-powered jet is more ambiguous, but recent works that have considered jet precession in SGRBs assumed that a jet powered by the BZ mechanism will align with the disk angular momentum vector \citep{RRS, LWG} because the magnetic field is anchored in the disk.  This assumption has been further supported by NR simulations of force-free electromagnetic fields around spinning BHs \citep{Palenzuela+10}, which found that the direction of Poynting flux from the BZ mechanism is governed by larger-scale magnetic fields and not the BH spin vector, although we note that GRMHD simulations of tilted accretion flows with matter have until very recently been unable to resolve jets \citep{Fragile08}.  While this paper was under peer review, the first GRMHD results on tilted jets appeared in the literature \citep{MTB13}; in these simulations a collimated jet aligned with the BH spin vector out to a distance of $\sim 100 R_{\rm g}$ before bending to align with the disk angular momentum vector.  Thus radiation from BZ-powered jets may contain precessing and nonprecessing components, depending on the distance from the BH where the radiation originates.

A final self-consistency check on our model involves the expected duration of significant jet luminosity from these events.  Although the details are uncertain, a simple semianalytic model for jet luminosity \citep{KPK12} found that accretion rates above $\dot{M}_{\rm d} \gtrsim 0.003-0.01 M_{\odot}~{\rm s}^{-1}$ are required to sustain a luminous GRB (this model also found that the BZ mechanism typically dominates $\nu\bar{\nu}$ annihilation).  

From this criterion and Eq. \eqref{MDotMetz}, we see that our fiducial SGRB duration of $\approx 1~{\rm s}$, originally chosen on observational grounds \citep{kmf+93, Nakar2007}, is well motivated internally as well.  This criterion for jet duration implies that the event typically lasts between one and several viscous time scales, which are largely controlled by $\alpha$.  For $\alpha=0.01~(0.1)$, the initial viscous time scale for a $8 M_{\odot}$ BH is $t_{\rm visc, 0} \approx 0.7~{\rm s}~(0.07~{\rm s})$.  This time scale grows as the disk expands, so that an SGRB that stays active for $1~{\rm s}$ lasts for $\lesssim 1.5~(15)$ viscous time scales.  As only the lower viscosities we consider are likely to produce $N_{\rm cycles} \gtrsim 1$, a typical BH-NS SGRB accompanied by an observably precessing jet will last for a few viscous times.  When we calculate $N_{\rm cycles}$ at later points in this paper, we integrate between times of 20 ms and 1s.  The lower bound is the approximate time by which simulations of misaligned BH-NS merger disks reach a steady state and begin exhibiting solid body precession \citep{FDKT}.

\section{Progenitor Binaries}
Because large amplitude precession requires large amplitude misalignment of the postmerger BH  and its accretion disk, we must consider which CO mergers can actually produce misaligned disks.  Despite their larger premerger spin-orbit misalignment, NS-NS mergers are unlikely to produce significantly misaligned disks.  If one or both members of the NS-NS binary were millisecond pulsars, disk precession could be feasible: spin angular momentum $J_{\rm MSP}\approx \frac{2}{5}M_{\rm NS}R_{\rm NS}^2\Omega_{\rm NS}\approx 1\times 10^{42} {\rm kg~m^2/s}$, and the orbital angular momentum at the disruption radius $L_{\rm NS-NS}\approx 2M_{\rm NS}R_{\rm NS}^2\sqrt{2GM_{\rm NS}/R_{\rm NS}^3}\approx 1\times 10^{43}{\rm kg~m^2/s}$.  Thus, the misalignment angle $\psi_{\rm d}$ between the postmerger accretion disks (which we assume to lie in the initial orbital plane) and the BH spin axis will be $\sim 5^{\circ}$ in NS-NS mergers involving one star with a ms spin period, if we assume initially orthogonal spin and orbital angular momentum vectors.  Standard population synthesis channels indicate, however, that the one recycled component of NS binaries typically has a minimum spin period of $\approx 4~{\rm ms}$ \citep{Willems+08}, which would imply $\psi_{\rm d} \lesssim 1^{\circ}$, a value that is likely too small to carry significant observational consequences.  Although the current sample of NS-NS binary spin measurements is limited, the fastest rotator discovered so far has a 22 ms spin period \citep{BDP03}, far too slow to produce large amplitude disk precession.

In BH-NS mergers, however, the BH may possess a larger natal reservoir of spin angular momentum, allowing for greater misalignment between the postmerger BH and the disk formed from NS debris (which we have assumed to lie in the initial orbital plane).  Natal spin is the most relevant quantity, although subsequent mass transfer onto the BH may produce modest changes in $a_{\rm BH}$ \citep{Belczynski+08}.  For a BH-NS system, the relevant numbers are $J_{\rm BH}= a_{\rm BH}GM_{\rm BH}/c$ and $L_{\rm BH-NS}\sim (1+q)M_{\rm BH}R_{\rm NS}^2\sqrt{(1+q)GM_{\rm BH}/R_{\rm NS}^3}$, where $a_{\rm BH}$ is the dimensionless black hole spin, and we define the mass ratio $q=M_{\rm NS}/M_{\rm BH}$.  For $M_{\rm BH}=5M_{\odot} (10M_{\odot})$ and $a_{\rm BH}=0.9$, the disk misalignment angle $\psi_{\rm d} \lesssim 20^{\circ} (30^{\circ})$, which is large enough to be observationally interesting.  Here we have also assumed initially orthogonal spin and orbital angular momentum vectors.
  
These simple Newtonian estimates motivate an investigation of BH-disk misalignment in BH-NS mergers, but are insufficient for accurately estimating either postmerger BH spin ${\bf a}_{\rm BH}'$ or the misalignment angle $\psi_{\rm d}$ between ${\bf a}_{\rm BH}'$ and ${\bf L}_{\rm BH-NS}$.  This is because they do not account for the fully dynamical, strong field GR effects that accompany compact object mergers.  Recently, empirical post-Newtonian (PN) formulas were derived to calculate these quantities in the case of BH-BH mergers \citep{Lousto10}.  These formulas have a large number of free parameters that were calibrated based on a suite of NR simulations.  Because disruptions of NSs by BHs are so marginal (i.e. occur so close to the ISSO), it is reasonable to expect these formulas to have some utility in making predictions for BH-NS mergers; we show in Appendix B that they are in fact quite accurate when tested against NR simulations of BH-NS coalescence.  For this reason, we use these PN formulas to calculate the final spin ${\bf a}_{\rm BH}'$ for a BH formed by the merger of a BH-NS with mass ratio $q$, premerger BH spin ${\bf a}_{\rm BH}$, and a premerger spin-orbit misalignment angle $\psi$.  We approximate the premerger NS spin magnitude as ${\bf a}_{\rm NS}=0$.  Finally, we calculate $\cos\psi_{\rm d}={\bf a}_{\rm BH}' \cdot {\bf L}_{\rm BH-NS}/(a_{\rm BH}'L_{\rm BH-NS})$.

Even though the PN formulas in Appendix B reproduce ${\bf a}_{\rm BH}'$ with reasonable accuracy, our assumption that the postmerger disk tilt $\psi_{\rm d}$ is described by the angle between the initial orbital plane and ${\bf a}_{\rm BH}'$ may overestimate $\psi_{\rm d}$.  When we compare our calculations to the late-time ($40~{\rm ms}$) results of Ref. \citep{FDKT}, we appear to overestimate $\psi_{\rm d}$ by a factor $\sim 3$, although our approach provides a significantly more accurate estimate of early-time disk tilt.  Similar evolution of disk tilt is not seen in steady-state GRMHD disk simulations \citep{FBAS} but is likely at least partially physical for the early stages of BH-NS mergers, as the disk adjusts to an equilibrium configuration.  However, numerical viscosity may also play a role in reducing the disk tilt in misaligned NR simulations \citep{FFComm}.  For simplicity, we do not model the time evolution of $\psi_{\rm d}$, but this probably results in overestimates of $\approx 2$, which we note again in Sec. V.

The final component in our calculation is a criterion for SGRB production in a BH-NS merger.  The tidal radius, defined in the Newtonian limit as $r_{\rm t}= R_{\rm NS}q^{-1/3}$, is appealing for this purpose:  only a fraction $f_{\rm GRB}$ of all BH-NS mergers will produce an accretion disk and jet, because if $r_{\rm t}<r_{\rm ISSO}$, the NS is swallowed whole.  However, while the tidal radius is cleanly defined in other contexts \citep{Rees88}, in the case of relativistic, comparable mass mergers it is not obvious that the Newtonian definition is applicable \citep{Fishbone73}.  Furthermore, the gravitational radius $r_{\rm g}\sim R_{\rm NS}$.  In light of these complications, we adopt a fitting formula for the initial remnant disk mass $M_{\rm dis}$, calibrated from NR simulations of aligned BH-NS mergers \citep{Foucart12}:
\begin{equation}
\frac{M_{\rm dis}}{M_{\rm NS}}=0.415q^{-1/3}\left(1-2\frac{GM_{\rm NS}}{c^2R_{\rm NS}}\right)-0.148\frac{r_{\rm ISCO}}{R_{\rm NS}}. \label{MdFF}
\end{equation}
Although this fitting formula was calibrated from NR data on aligned mergers, we generalize it to misaligned mergers by substituting $r_{\rm ISSO}$ for $r_{\rm ISCO}$.  This appears to reproduce NR simulations of misaligned postmerger disks reasonably well (Appendix B).  Overall, the fitting formula's accuracy is comparable to a more complex theoretical model for $M_{\rm dis}$ using the ``affine ellipsoids'' approximation \citep{PTR11}.

The limiting value of $M_{\rm dis}/M_{\rm NS}$ required to produce a SGRB is highly uncertain, but past theoretical work assuming jets are powered by $\nu \bar{\nu}$ annihilation has suggested that SGRBs are viable for $M_{\rm dis}/M_{\rm NS}>0.01$ \citep{RuffertJanka99}; likewise, a recent attempt to observationally infer disk masses assuming $\nu\bar{\nu}$ annhilation \citep{FanWei11} found $0.01<M_{\rm dis}/M_{\rm NS}<0.1$.  As mentioned in Sec. II, more recent work examining the BZ mechanism suggested that initial accretion rates of $\dot{M}_{\rm d} \gtrsim 0.003-0.01~M_{\odot}~{\rm s}^{-1}$ are necessary to power a SGRB, which in combination with Eq. \eqref{MDotMetz} implies that a SGRB 1 second in duration needs $M_{\rm dis} \gtrsim 0.02-0.05~M_{\odot}$.  In this work, we take $M_{\rm dis}/M_{\rm NS}>0.05$ as the cutoff for SGRB production, but we discuss the effects of stricter and weaker criteria in Sec. IV and V.

\section{Distributions}
We now integrate the above analytic criteria over distributions of progenitor masses and spins to find distributions of $f_{\rm GRB}$, $\psi_{\rm d}$, $N_{\rm cycles}$, and $T_{\rm prec}(t_{1/2})$.  Because the distributions of progenitor quantities are not at present precisely constrained by observation or population synthesis, we consider a wide range of possibilities to bracket the available parameter space.  

For our fiducial case, we take the parametric BH mass function from Ref. \citep{OPNM} (hereafter the ``OPNM mass function''), given by
\begin{equation}
P_{\rm OPNM}(M_{\rm BH})=\frac{e^{\frac{M_{\rm c}}{M_{\rm scale}}}}{M_{\rm scale}}
\begin{cases}
e^{-\frac{M_{\rm BH}}{M_{\rm scale}}}, &M_{\rm BH}>M_{\rm c} \\
0, &M_{\rm BH} \le M_{\rm c},
\end{cases}
\end{equation}
where the best-fit values were found to be $M_{\rm scale}\approx 1.7 M_{\odot}$ and $M_{\rm c}\approx 6.2M_{\odot}$ \citep{OPNM}.  An important qualitative feature of the OPNM mass function is the large mass gap between NSs and the lowest-mass BH.  Although the mass gap has been known from observations for some time \citep{BJC98}, it is not fully understood theoretically \citep{FK99}.  Recent population synthesis efforts have had some success in reproducing it \citep{BWF12} provided strong assumptions are made about the growth of instabilities in supernova explosion mechanisms.  Motivated by recent observations \citep{KBFK} suggesting that the mass gap may be less distinct than in Ref. \citep{OPNM}, we consider as an alternate case a Gaussian mass function where the best-fit values for mean mass $\mu_{\rm BH}$ and dispersion $\sigma_{\rm BH}$ were found to be $7.35M_{\odot}$ and $1.25 M_{\odot}$, respectively \citep{Farr11}.

Recent observations have measured spins for 7 stellar mass BHs and placed upper or lower limits on spins for 3 more \citep{McClintock11, GMR11, SMR12}.  Although observations of more systems are needed, the current spin distribution is noticeably bimodal.  Because of the small number of data points we do not attempt to fit a parametrized spin function, and instead simply take a flat prior on $a_{\rm BH}$, sampling it uniformly in the ranges $(0, 0.3)$ and $(0.7, 1)$ for our fiducial, ``bimodal'' case.  For nonfiducial cases, we also consider three alternate spin functions.  The ``flat,'' ``slow,'' and ``fast'' cases uniformly sample $a_{\rm BH}$ along the intervals $(0, 1)$, $(0, 0.5)$, and $(0.5, 1)$, respectively.  We explore a variety of possible spin distributions because theoretical expectations are quite uncertain; in particular, the spin distribution at time of merger depends strongly on both the (unknown) birth spins, and on poorly constrained details of a hypercritical common envelope phase \citep{OKK05, Belczynski+08}.

Kicks resulting from asymmetric supernova explosions are expected to produce spin-orbit misalignment in BH-NS binaries.  Past research has constrained the allowed premerger misalignment angle $\psi$ as a function of progenitor masses and separation and kick velocity distributions \citep{Kalogera00}.  More recent population synthesis of BH-NS binaries has found a wide spread in premerger spin-orbit misalignment $\psi$, but with $\sim 50\%$ of systems possessing $\psi<45^{\circ}$ \citep{Belczynski+08}.  Our fiducial, ``prograde'' case samples the premerger spin-orbit misalignment uniformly in $\psi$ between $0^{\circ}$ and $90^{\circ}$,  but we also consider an alternate, ``isotropic'' case where $\psi$ is sampled uniformly from $0^{\circ}$ to $180^{\circ}$; physically this would represent larger supernova kicks, or formation of binaries through nonstandard processes such as dynamical capture.

Finally, we sample NS masses from a Gaussian distribution peaked at a mean $\mu_{\rm NS}=1.35M_{\odot}$ with standard deviation $\sigma_{\rm NS}=0.13 M_{\odot}$.  These values, taken from the double NS binaries examined in Ref. \citep{KKT}, are in good qualitative agreement with other studies of the NS mass function \citep{VRH}.  Because most NS equations of state that are not in conflict with observations of $\approx 2M_{\odot}$ NSs \citep{Dem10} are roughly constant radius in the relevant mass range, we take a fiducial radius of $13.5~{\rm km}$, but as an alternate case consider a NS radius of $11~{\rm km}$.  For reasons described earlier, we neglect NS spin.

With these distributions defined, we are now ready to populate a large Monte Carlo sample of BH-NS mergers.  Our precise procedure is as follows, for any desired set of distributions: 
\begin{enumerate}
\item Generate $2\times 10^5$ BH masses $M_{\rm BH}$, spin magnitudes $a_{\rm BH}$, initial misalignment angles $\psi$, and NS masses $M_{\rm NS}$.
\item Compute the premerger $r_{\rm ISSO}$ from Eq. \eqref{ISSOGeneral}.
\item Calculate the postmerger BH mass $M_{\rm BH}'$, BH spin $a_{\rm BH}'$, and spin-disk misalignment $\psi_{\rm d}$ using Eqs. \eqref{deltaM} and \eqref{aPrime}.
\item Flag the disruption as GRB producing if $M_{\rm dis}/M_{\rm NS}>0.05$.
\item Compute $N_{\rm cycles}$, $T_{\rm prec}(t_{1/2})$, and $f_{\rm GRB}$ (using the postmerger $r_{\rm ISSO}'$).  In these calculations, set $R_{\rm dis}=r_{\rm ISSO}'$.
\end{enumerate}

\begin{table*}
\label{Scenarios}
\centering
\begin{tabular}{ l || c | c | c | c | c | c | c | c | c}
  Scenario & BH Masses & BH Spins & $\psi$ & $M_{\rm dis}/M_{\rm NS}$ & $R_{\rm NS}$ & $f_{\rm GRB}$ & $\langle N_{\rm cycles}\rangle$ & $\langle T_{\rm prec}(t_{\rm 1/2}) \rangle$ & $\langle \psi_{\rm d} \rangle$\\
  \hline                        
  A & OPNM & bimodal & prograde & 0.05 & 13.5 km & 0.303 & $6.0~(16.7, 1.6)$ & $50~(21, 180)~{\rm ms}$ & $20.9^{\circ}$\\
  B & Gaussian & bimodal & prograde & 0.05 & 13.5 km & 0.333 & $5.7~(16.1, 1.5)$ & $53~(22, 188)~{\rm ms}$ & $20.9^{\circ}$\\
  C & OPNM & bimodal & prograde & 0.05 & 11 km & 0.131 & $7.0~(20.3, 1.8)$ & $40~(15, 143)~{\rm ms}$ & $15.9^{\circ}$\\
  D & OPNM & flat & prograde & 0.05 & 13.5 km & 0.229 & $5.3~(14.9, 1.4)$ & $57~(24, 198)~{\rm ms}$ & $19.0^{\circ}$\\
  E & OPNM & slow & prograde & 0.05 & 13.5 km & 0.010 & $2.4~(6.3, 0.7)$ & $116~(53, 376)~{\rm ms}$ & $9.3^{\circ}$\\
  F & OPNM & fast & prograde & 0.05 & 13.5 km & 0.447 & $5.4~(15.2, 1.5)$ & $56~(23, 194)~{\rm ms}$ & $19.3^{\circ}$\\
  G & OPNM & bimodal & isotropic & 0.05 & 13.5 km & 0.248 & $4.5~(12.5, 1.2)$ & $108~(58, 364)~{\rm ms}$ & $32.0^{\circ}$\\
  H & OPNM & bimodal & isotropic & 0.05 & 11 km & 0.099 & $6.4~(18.3, 1.7)$ & $56~(34, 239)~{\rm ms}$ & $23.2^{\circ}$\\
  I & OPNM & bimodal & prograde & 0.01 & 13.5 km & 0.348 & $5.7~(15.9, 1.5)$ & $54~(23, 190)~{\rm ms}$ & $22.6^{\circ}$\\
  J & OPNM & bimodal & prograde & 0.2 & 13.5 km & 0.102 & $7.4~(21.5, 1.9)$ & $37~(14, 136)~{\rm ms}$ & $15.3^{\circ}$\\
\end{tabular}
\caption{Choices of black hole mass, black hole spin, and spin-orbit misalignment distributions we have surveyed in this paper.  We also summarize key results here: the fraction of BH-NS mergers that can produce a disk and GRB ($f_{\rm GRB}$), a typical precession period in milliseconds $\langle T_{\rm prec}(t_{1/2}) \rangle$, and the mean postmerger misalignment angle between the disk and the BH equatorial plane $\langle \psi_{\rm d}\rangle$.  The first number printed in the $\langle T_{\rm prec} \rangle$ and $\langle N_{\rm cycles} \rangle$ columns is the fiducial value for $\alpha=0.03$, while the numbers in parentheses represent $\alpha=0.01$ and $\alpha=0.1$ cases.}  
\end{table*}

\begin{figure*}[!t]
\includegraphics[width=180mm]{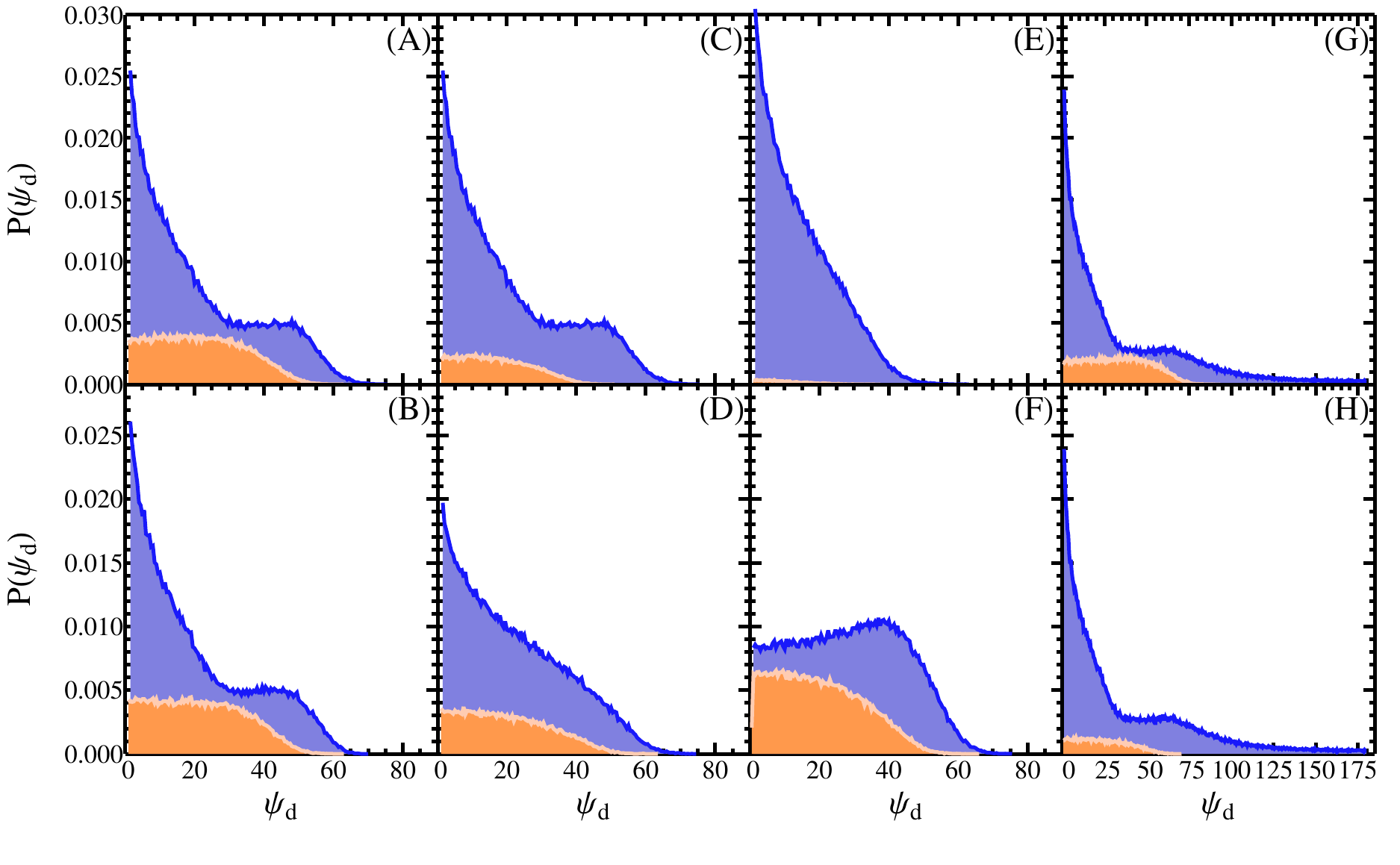}
\caption{The probability distribution of postmerger misalignment angles $\psi_{\rm d}$ for scenarios A-H (in respectively labeled panels).  The vertical axis is a probability and the horizontal axis is $\psi_{\rm d}$.  The dark blue curves show the $\psi_{\rm d}$ distribution for all $2\times 10^5$ BH-NS mergers in each scenario, while the smaller, light orange curves show only the subset of events where the NS is disrupted outside the ISSO (i.e. the subset where a disk and jet can form).  Generally, $f_{\rm GRB}$, the ratio of the light orange area to the dark blue area, falls for bottom-heavy spin distributions and smaller NS radii.  All distributions of $\psi_{\rm d}$ cut off fairly sharply above $50^{\circ}$.} \label{psidFig}
\label{psidFig}
\end{figure*}

\begin{figure}[!t]
\includegraphics[width=85mm]{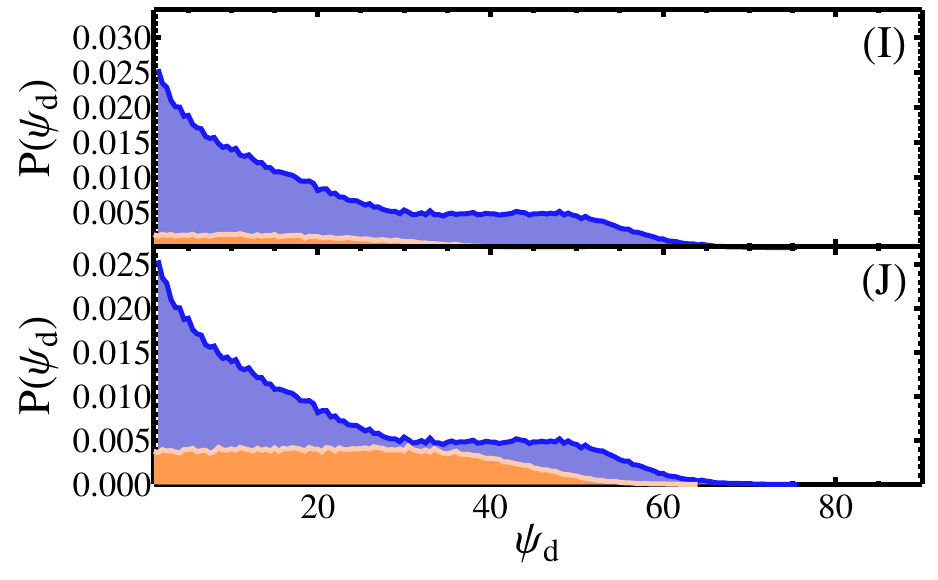}
\caption{The probability distribution of postmerger misalignment angles $\psi_{\rm d}$ for scenarios I and J, illustrating the dependence of $f_{\rm GRB}$ and $\psi_{\rm d}$ on the strictness of our requirement for $M_{\rm dis}/M_{\rm NS}$.  Specifically, in the top panel we have imposed the strict requirement that $M_{\rm dis}/M_{\rm NS}>0.2$ in order to produce a SGRB, while in the lower panel we have imposed the much laxer requirement of $M_{\rm dis}/M_{\rm NS}>0.01$.  In all other respects both of these cases are identical to scenario A.  Results in the bottom panel  are quite similar to scenario A, but in the top panel, $f_{\rm GRB}$ has been strongly suppressed, particularly at higher $\psi_{\rm d}$.  The axes and curves are the same as in Fig. \ref{psidFig}.}
\label{psidFigLaxStrict}
\end{figure}

\begin{figure*}[!t]
\includegraphics[width=180mm]{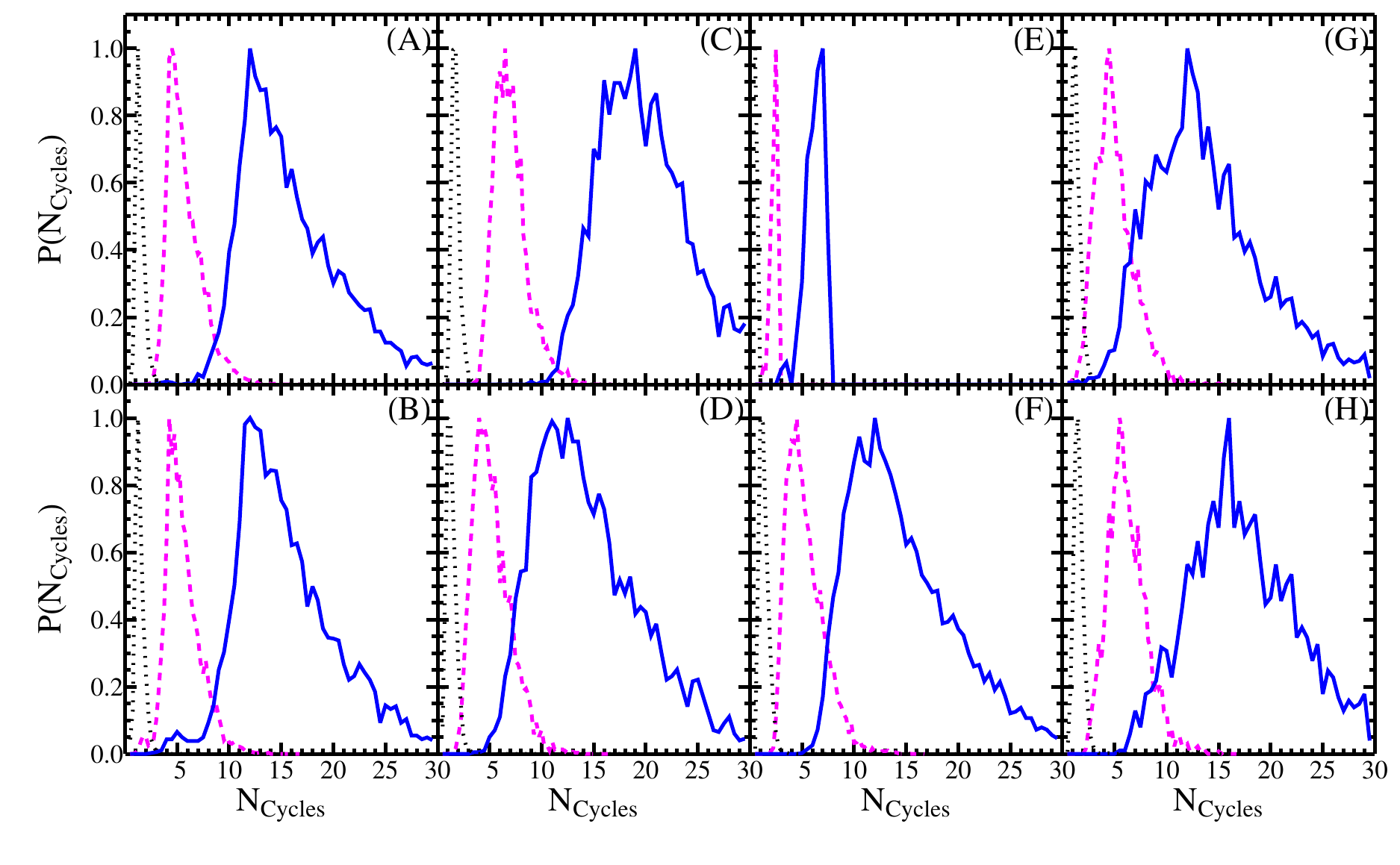}
\caption{Probability distributions of $N_{\rm cycles}$ for scenarios A-H.  As in Fig. 1, the dotted black lines represent $\alpha=0.1$, the dashed magenta lines $\alpha=0.03$, and the solid blue lines $\alpha=0.01$.  There is relatively little variation between progenitor scenarios, with the notable exception of scenario E (slow spins).  The total number of precession cycles accumulated depends strongly on $\alpha$.}
\label{NCyclesFig}
\end{figure*}

\begin{figure*}[!t]
\includegraphics[width=180mm]{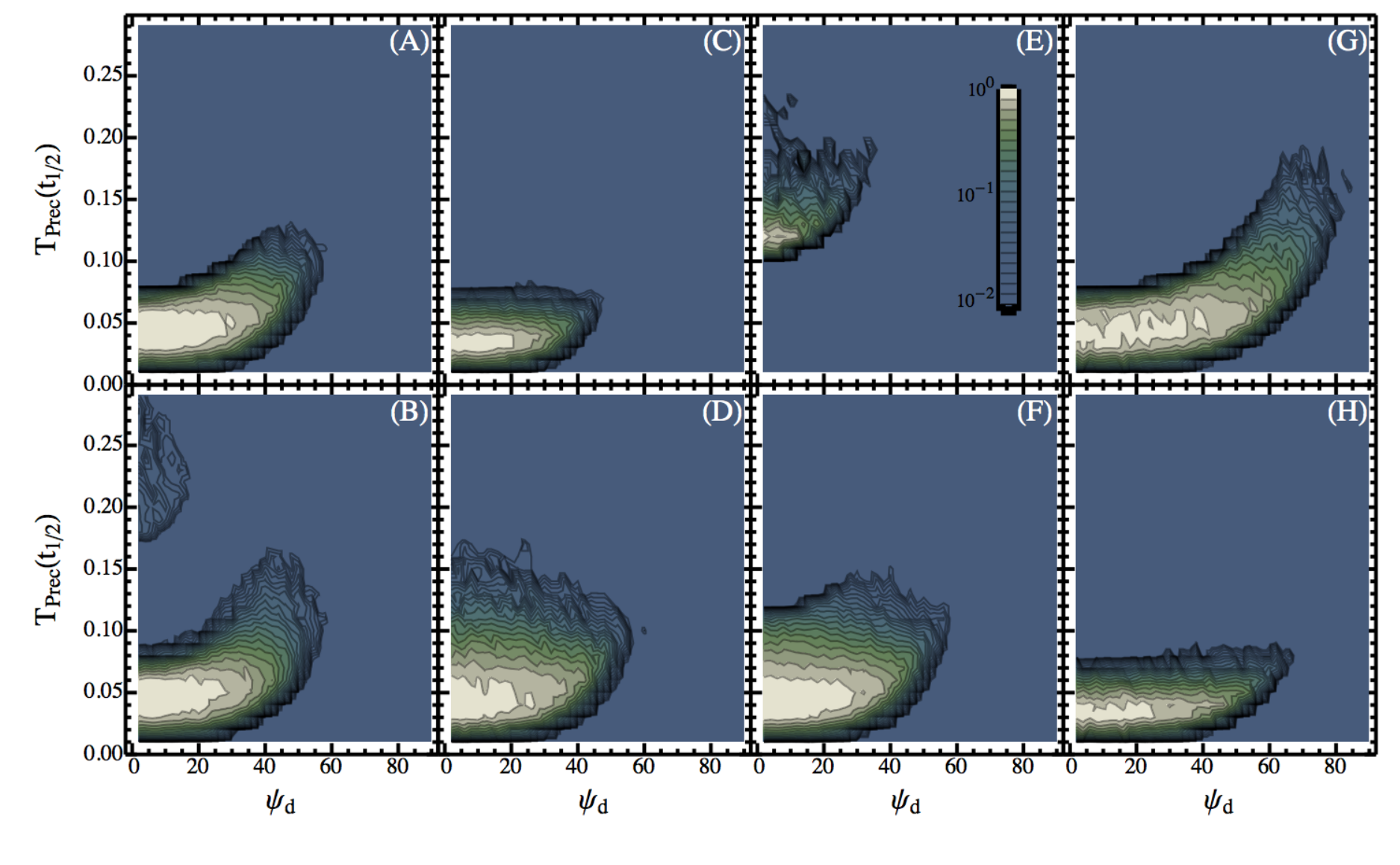}
\caption{Each contour plot in this figure shows the range of ``halfway'' precession periods (vertical axis) and postmerger misalignment angles $\psi_{\rm d}$ (horizontal axis) for every BH-NS merger that produces a disk.  $T_{\rm prec}$ is plotted in seconds, and in all scenarios is generally $\le 0.4 {\rm s}$, although we note again that it will grow with time as the disk viscously spreads outward.  As in previous plots, we show scenarios A-H.  The logarithmic contour scale is included in panel (e); the range in colors covers a probability range of 2 orders of magnitude.  In this plot we use the fiducial value $\alpha=0.03$.}
\label{TprecFig}
\end{figure*}

In Table I, we summarize the scenarios considered in this work, along with averaged results.  Typical precession periods are $\langle T_{\rm prec}(t_{1/2}) \rangle \approx 20~{\rm ms}-200~{\rm ms}$, with mean disk-BH misalignment typically $\langle \psi_{\rm d} \rangle \approx 20^{\circ}$.  These results are generally insensitive to variation of assumptions about the progenitor population.  The fraction of BH-NS mergers that produce SGRBs ($f_{\rm GRB}$) varies more with the properties of progenitor binaries, and, in particular, depends strongly on the value of premerger BH spin $a_{\rm BH}$.  Values of $a_{\rm BH}\gtrsim 0.5$ are generally necessary to produce a postmerger disk.  Scenario A, our fiducial case, results in GRBs for $\approx 30\%$ of BH-NS mergers.  Results for scenario A are quite similar to those in scenarios B (Gaussian BH mass function) and D (flat prior on the BH spin distribution).  The probability of disruption of the NS falls by a factor $\approx 2$ in scenario C (softer NS equation of state), and becomes almost negligible in scenario E (bottom-heavy spin distribution).  The fraction of BH-NS mergers capable of producing SGRBs is maximized in scenario F (fast spins), reduced somewhat if we switch to an isotropic $\psi$ distribution (scenario G), and reduced significantly for a stricter lower cutoff on $M_{\rm dis}/M_{\rm NS}$ (scenario J).  Our fiducial scenario comes close to maximizing $f_{\rm GRB}$; only switching over to top-heavy BH spin distributions (scenario F) or laxer SGRB criteria (scenario I) increase its value.  More detail on $f_{\rm GRB}$ and distributions of $\psi_{\rm d}$ can be seen in Fig. \ref{psidFig}.

In Fig. \ref{psidFigLaxStrict}, we explore the sensitivity of $f_{\rm GRB}$ and $\psi_{\rm d}$ to our uncertain assumption about the minimum normalized disk mass, $M_{\rm dis}/M_{\rm NS}$, needed to produce a SGRB.  In our standard scenarios, we take a cutoff value of 0.05, but in scenarios I and J we change this cutoff to 0.01 and 0.2, respectively.  While relaxing the cutoff seems to have little effect on our overall results, increasing the cutoff above $M_{\rm dis}/M_{\rm NS}\approx 0.1$ very quickly suppresses $f_{\rm GRB}$, and biases those SGRBs that are produced toward slightly tilted accretion disks.

In general, $\psi_{\rm d}$ is strongly cut off above $40^{\circ}-50^{\circ}$ in all scenarios except E, and there is little variation between the mean value $\langle \psi_{\rm d} \rangle$ in each of our progenitor scenarios.  Specifically, $\langle \psi_{\rm d} \rangle$ ranges from $9^{\circ}$ to $32^{\circ}$.  Interestingly, our two scenarios with isotropic premerger $\psi$ only modestly extend the range of postmerger $\psi_{\rm d}$ values: strongly misaligned ($\psi>90^{\circ}$) BH-NS mergers simply do not produce massive disks.  As mentioned earlier, our calculation of $\psi_{\rm d}$ likely overestimates its true value by a factor $\approx 2$, because the tilt angle will decrease as the disk settles into equilibrium, so the true range of $\langle \psi_{\rm d} \rangle$ would more accurately be $\sim 5^{\circ}$ to $\sim15^{\circ}$.  For rapidly spinning BHs, it is possible that our $\psi_{\rm d}$ estimate accurately describes the first few precession cycles, before disk tilt has time to decrease.

As can be seen in Table I, the key observable $\langle N_{\rm cycles} \rangle$ depends much more sensitively on $\alpha$ than on the parameters of BH-NS binary populations.  As usual, the outlier is scenario E, which is biased toward few cycles and long precession time scales, but all other scenarios produce $4.5 \lesssim \langle  N_{\rm cycles} \rangle \lesssim 7.5$ and $30~{\rm ms} \lesssim \langle T_{\rm prec}(t_{1/2}) \rangle \lesssim 100~{\rm ms}$ for our fiducial $\alpha=0.03$ case.  We plot distributions of $N_{\rm cycles}$ in Fig. \ref{NCyclesFig}, where we see that low $\alpha$ values produce many more disk precession cycles.  

In Fig. \ref{TprecFig} we plot contours to indicate the relative probability of BH-NS mergers producing combinations of $T_{\rm prec}$ and $\psi_{\rm d}$.  These two quantities in general appear fairly uncorrelated, although in some scenarios (A, B, and G) we do see a weak positive correlation, indicating that the most dramatically precessing SGRB disks will typical precess with longer periods.  In no scenario do we see typical $T_{\rm prec}(t_{1/2})$ values above 0.4 s; if we set aside scenario E, only a small fraction of events have $T_{\rm prec}(t_{1/2})$ above 0.1 s when $\alpha=0.03$.  Scenarios C and H (soft NS equation of state) generally have the shortest precession periods, as a combination of rapidly spinning BHs and fairly low $\psi$ are required to disrupt these more compact NSs.  Generally,  $T_{\rm prec}(t_{1/2}) \gtrsim 0.01$ s in all scenarios.

\section{Discussion}
In this paper, we have explored much of the relevant parameter space for BH-NS mergers and found that quantities relevant for disk precession are generally insensitive to the assumptions we have made about progenitor binary parameters.  The important results of our calculations ($ \langle T_{\rm prec}(t_{1/2}) \rangle$, $\langle N_{\rm cycles} \rangle$, $ \langle \psi_{\rm d} \rangle$, $f_{\rm GRB}$) typically change by factors $\lesssim 2$ as we have varied our assumptions about the underlying populations of premerger BH-NS binaries.  The one exception to this concerns premerger BH spin: in order for a significant fraction of BH-NS mergers to produce SGRBs, $a_{\rm BH}\gtrsim 0.5$ is required.  Because of the difficulty of increasing BH spin through premerger mass transfer \citep{Belczynski+08}, this is equivalent to requiring modestly large natal $a_{\rm BH}$ values.  SGRB formation could also be inhibited if the typical NS radius is significantly below the smallest value we consider (11 km).  The strongest precession effects (i.e. large misalignment $\psi_{\rm d}$ and short period $T_{\rm prec}$) arise from populations with top-heavy BH spin distributions and stiff NS equations of state.  Our results are notably insensitive to the choice of BH mass function, so long as the mass function peaks near $6 M_{\odot}$, as is suggested by observations of x-ray binaries.  

On the other hand, our results do depend fairly sensitively on the details of how the disk will viscously spread outward.  Using the $\alpha$ viscosity parametrization, we have seen that effective $\alpha$ values $\gtrsim 0.1$ will strongly suppress the number of observable precession cycles, to the point where jet precession will rarely if ever be detectable.  Likewise, $\alpha \gtrsim 0.1$ could be large enough to induce a Bardeen-Petterson warp in the thinner, inner regions of the disk, limiting global precession.  We note again, however, that what matters most for the time evolution of $T_{\rm prec}$ is not the value of $\alpha$ in the inner regions of the disk, where GR effects enhance the turbulent viscosity produced by the MRI \citep{McKinneyNarayan07}, but rather the outer regions of the disk, where local shearing box simulations that find lower ($0.01 \lesssim \alpha \lesssim 0.1$) levels of MRI-generated viscosity are more appropriate \citep{DSP10}.

If jets align with the disk angular momentum axis, then they will precess around the total angular momentum vector by an angle $\approx \psi_{\rm d}$, because $J_{\rm BH}$ is significantly larger than $J_{\rm disk}$.  In this case, observations of SGRBs associated with BH-NSs will often be marked by a clear ``lighthouse effect,'' so long as $\psi_{\rm d} \gtrsim \theta_{\rm jet}$.  This seems plausible, as observations of jet breaks in SGRBs suggest typical opening angles of $\sim 10^\circ$ \citep{sbk+06}.  If $\psi_{\rm d} \lesssim \theta_{\rm jet}$, then jet precession would, typically, be encoded more subtly as a variation in the portion of the jet presented to the observer.  Alternatively, if jets align with the BH spin axis, they will precess by a much smaller amount, since the angle between the BH spin vector and the total system angular momentum vector is $\cos\psi_{\rm BH}=(1+j\cos\psi_{\rm d})/\sqrt{1+2j\cos\psi_{\rm d} + j^2}$, where $j=J_{\rm d}/J_{\rm BH}$.  This angle will initially be smaller than $\psi_{\rm d}$ by a factor $\sim 10$ if  $J_{\rm d} \sim 0.1J_{\rm BH}$ \citep{FDKT}, and this difference will increase slowly in time as angular momentum is accreted from the disk onto the BH (or quickly, if angular momentum is lost in outflows).  Unless the typical $\theta_{\rm jet}$ is quite small, less than a few degrees, the dramatic lighthousing effect previously discussed would be unlikely.  A quasiperiodic signal will still be present in all these scenarios, but it will be strongest for tightly collimated jets aligned with ${\bf J}_{\rm disk}$.  When we correct our results for the $\approx 2$ overestimate in $\psi_{\rm d}$, we have found $\langle \psi_{\rm d}\rangle \sim \theta_{\rm jet}$, although there are substantial observational uncertainties regarding the distribution of jet opening angles.

Detection of a precession signal appears observationally feasible: our study found a floor of $\sim 25$ ms for the ``halfway'' precession period $\langle T_{\rm prec}(t_{1/2})\rangle$, but precession periods of $\sim 50$ msec were much more common, and in all cases the precession period grows in time.  Both the Swift satellite's Burst Alert Telescope and the Fermi satellite's Gamma-ray Burst Monitor have much finer intrinsic timing resolution, so the observability of pure precession signals will be ultimately limited by photon-counting statistics.  Analysis of Fermi SGRB light curves finds evidence of structure in the $\sim 10-50$ msec range \citep{Bhat+12}, while Swift's Burst Alert Telescope also produces signals with structure in the tens of msec \citep{Rowlinson+10}.  For comparison, the BATSE archival data sample contains sufficiently high-resolution timing data to enable a search for $\sim$ kHz frequency QPOs \citep{PBM, KLW}, indicating that the much longer precession signals should be clearly resolvable.

On the other hand, precession is unlikely to be the only source of variability in the noisy prompt emission light curves of SGRBs.  Observational constraints on the nature and sources of SGRB variability are limited by the low signal-to-noise ratio and small rate of these events (e.g., SGRBs made up $\approx 1/4$ of the BATSE sample).  However, analysis of BATSE data indicates that SGRB prompt emission is highly variable, with typical variability time scales $\sim 50-100~{\rm ms}$, uncorrelated with the total burst duration \citep{NP02}.  In rare cases, variability on millisecond or shorter time scales has been seen \citep{SNB98}.  From a theoretical perspective, the observed variability is generally attributed to internal rather than external shocks \citep{SP97}, although relativistic turbulence \citep{LNP09} and magnetic dissipation in Poynting-dominated outflows \citep{Thompson06} are other possibilities.  All of these would represent sources of confusion for a realistic precession signal, although a lighthouse effect for high-amplitude precession could serve as a distinguishing feature.

The two largest challenges for observability are (1) the relatively low number of precession cycles, and (2) the rapidly growing precession period $T_{\rm prec}$.  However, we have shown that for most progenitor populations there exists a large value tail to the $N_{\rm cycles}$ distribution, from which events with $N_{\rm cycles}\gtrsim 5$ can be observed, provided $\alpha < 0.1$.  Furthermore, searches for evolving quasiperiodicity are feasible, provided the time axis of time series data can be rescaled to match appropriate theoretical models.  Past searches for simple periodicity in SGRBs \citep{KLW} would not have been able to detect the more complex time evolution of a realistic jet precession signal.  In this paper we have employed a simple analytic model for the viscous spreading of the disk, and for the first time have found that precession periods grow as $t^{4/3}$.  We hope that this will provide a starting point for searches for jet precession in SGRBs, but detailed hydrodynamical, and perhaps GRMHD, simulations are necessary to validate or refine this analytic expectation.  We also note that both of these challenges are mitigated if a significant fraction of disk angular momentum is lost through an outflow, which will reduce the late-time scaling of $T_{\rm prec}$ to $t^{4/5}$ - although in this scenario, only emission aligned with $\bf{J}_{\rm d}$, not $\bf{J}_{\rm BH}$, will precess significantly.

Pulsation of prompt emission may not be the only observable implication of disk precession in SGRBs.  A precessing jet will sweep out a larger solid angle in the sky, enabling BH-NS mergers to make up a larger fraction of the observed SGRB rate than would be implied by a simple calculation (i.e. the intrinsic BH-NS merger rate multiplied by $f_{\rm GRB}$).  This would likewise enhance observability of BH-NS optical afterglows \citep{BergerMetzger12}.

Unless there exists a large population of NS-NS binaries with millisecond spin periods (not accounted for by current observations or population synthesis estimates), SGRBs due to BH-NS mergers will be distinguishable from those due to NS-NS mergers by the presence of a quasiperiodic signal, with a typical period of $\sim 30-100$ ms.  We have shown that this signal is robust to a large number of assumptions about the progenitor binaries, but evolves quickly and could become difficult to observe if jets align with the BH spin axis, or if the postmerger disk viscosity is large.  A better understanding of both viscous disk spreading, and how GRB jet intensity varies with angle of observation, will aid future searches for this discriminant between SGRB progenitor binaries.

\begin{acknowledgments}
We would like to thank Francois Foucart, Tassos Fragos, Vassiliki Kalogera, Raffaella Margutti, Brian Metzger, Cole Miller, and Gabriel Perez-Giz for helpful discussions and suggestions.  N.S. and A.L. were supported in part by NSF Grant No. AST-0907890 and NASA Grants No. NNX08AL43G and No. NNA09DB30A.  E.B. acknowledges support for this work from the National Science Foundation under Grant No. AST-1107973, and by NASA/Swift AO7 Grant No. NNX12AD71G.
\end{acknowledgments}

\appendix

\section{Innermost Stable Spherical Orbits}
We follow the simple formalism of Perez-Giz 2013 to calculate the ISSO radius for Kerr metric geodesics and present it here for completeness (in this appendix we use geometrized units, G=c=1, for radial distance $r$ and BH spin $a$).  As mentioned in Sec. II, the ISSO is the innermost stable orbit at constant radius but fixed nonzero inclination around a Kerr BH.  We define an inclination angle $\iota$ such that $C=\cos\iota$, the Carter constant $Q=L^2\sin^2\iota$, and conserved vertical angular momentum $L_{\rm z}=LC$, i.e. $\iota=0$ corresponds to equatorial prograde orbits.  Calculation of the equatorial plane ISCO, $r_{\rm ISCO}$, is well documented in the literature \citep{BPT} and consists of finding the roots of the polynomial
\begin{equation}
Z( r)=(r(r-6))^2-a^2(2r(3r+14)-9a^2)=0,
\end{equation}  
with one root the prograde and one the retrograde ISCO.  The polar ($\iota=\pm\pi/2$) ISSO can be found at the root of
\begin{align}
P( r)=&r^3(r^2(r-6)+a^2(3r+4))\\
&+a^4(3r(r-2)+a^2)=0 \notag
\end{align}
that lies between $r=6$ and $r=1+\sqrt{3}+\sqrt{3+2\sqrt{3}}$.  Finally, the generic ISSO is a root of the polynomial
\begin{align}
S( r)=r^8Z( r)+&a^2(1-C^2)(a^2(1-C^2)Y( r) \label{ISSOGeneral} \\
&-2r^4 X(r )) \notag,
\end{align}
with auxiliary functions defined as
\begin{align}
X( r)=&a^2(a^2(3a^2+4r(2r-3)) \\
&+r^2(15r(r-4)+28))-6r^4(r^2-4)\notag  \\ 
Y( r)=&a^4(a^4+r^2(7r(3r-4)+36))+6r(r-2)\\
&\times(a^6+2r^3(a^2(3r+2)+3r^2(r-2))) \notag.
\end{align}
Specifically, $r_{\rm ISSO}$ is the root of $S(r )$ located between the appropriate $r_{\rm ISCO}$ (prograde or retrograde) and the polar ISSO.

\section{Post-Newtonian Merger Treatment}

The Ref. \citep{Lousto10} formulas discussed in Sec. III are presented in this appendix, along with a discussion of their applicability to BH-NS mergers.  These formulas are calibrated to define the outcome of a BH-BH merger, although we apply them more generally to the case of BH-NS mergers with masses $m_{1}$ and $m_{2}$, and mass ratio $q=m_{1}/m_{2}\le 1$.  We define the symmetric mass ratio $\eta=q/(1+q)^2$, and denote the dimensionless CO spin vectors ${\bf a}_{\rm i}$ as having components ${a}_{\rm i}^{||}$ and ${a}_{\rm i}^{\perp}$, which are parallel to and perpendicular to the binary angular momentum, respectively.  We further define $\Theta_{\pm}$ as the angle made between the radial direction and the vector ${\bf \Delta}_{\pm}=(m_{1}+m_{2})(m_2{\bf a}_2\pm m_1 {\bf a}_1)$.  Then the postmerger remnant mass is found to decrease by a fraction
\begin{align}
\delta M/M = \eta \tilde{E}_{\rm ISCO}+E_2 \eta^2 + E_3 \eta^3 + \frac{\eta^2}{(1+q)^2}\\ 
\times \Bigl( E_{\rm S} ({a}_2^{||}+q^2 {a}_1^{||}) + E_{\Delta}(1-q)({a}_2^{||}-q{a}_1^{||}) \label{deltaM}\notag \\
 + E_{\rm A} |{\bf a}_2+q{\bf a}_1|^2+ E_{\rm B}|a^{\perp}_2+qa^{\perp}_1|^2 \notag \\
 \times (\cos^2(\Theta_+ - \Theta_2) + E_{\rm C}) +E_{\rm D} |{\bf a}_2-q{\bf a}_1|^2 \notag \\ 
 + E_{\rm E}|a^{\perp}_2-qa^{\perp}_1|^2 (\cos^2(\Theta_- - \Theta_3) + E_{\rm F}) \Bigr). \notag
\end{align}
Here the energy lost during the inspiral from infinity down to the plunge is fit as
\begin{align}
 \tilde{E}_{\rm ISCO} \approx 1-\frac{\sqrt{8}}{3} + 0.103803 \eta + \frac{1}{36\sqrt{3}(1+q)^2}\\
 \times \Bigl(q(1+2q)a^{||}_1+(2+q)a^{||}_2 \Bigr) -\frac{5}{324\sqrt{2}(1+q)^2}\notag \\
 \times \Bigl( {\bf a}^2_2 - 3(a_2^{||})^2  -2q({\bf a}_1 \cdot {\bf a}_2 - 3a_1^{||} a_2^{||}) \notag \\
 + q^2({\bf a}_1^2-3(a_1^{||})^2) \Bigr). \notag
 \end{align}
The final spin vector of the postmerger BH is 
\begin{align}
&{\bf a}'=\Bigl(1-\frac{\delta M}{M}\Bigr)^{-2} \Bigl[ \eta \tilde{\bf J}_{\rm ISCO} + (J_2\eta^2+J_3\eta^3)\hat{\bf n}_{||} \label{aPrime} \\ \notag
 &+ \frac{\eta^2}{(1+q)^2} \Bigl( (J_{\rm A} (a_2^{||}+q^2 a_1^{||}) + J_{\rm B}(1-q)(a_2^{||}-qa_1^{||})) \hat{\bf n}_{||} \\ \notag
 &+(1-q) |{\bf a}^{\perp}_2 - q{\bf a}_1^{\perp} | \sqrt{J_{\Delta}\cos(2(\Theta_{\Delta}-\Theta_4))+J_{\rm M} } \hat{\bf n}_{\perp} \\ \notag
& + | {\bf a}_2^{\perp} + q^2 {\bf a}_1^{\perp} | \sqrt{J_{\rm S} \cos(2(\Theta_{\rm S}-\Theta_5)) + J_{\rm MS}} \hat{\bf n}_{\perp} \Bigr) \Bigr],
\end{align}
where the angular momentum radiated during the inspiral from infinity to the plunge is
\begin{align}
\tilde{\bf J}_{\rm ISCO} \approx \Bigl[ 2\sqrt{3}-1.5255862 \eta - \frac{1}{9\sqrt{2}(1+q)^2}\\
 \times \Bigl( q(7+8q)a^{||}_1 + (8+7q) a_2^{||} \Bigr)+ \frac{2}{9\sqrt{3}(1+q)^2} \notag \\
 \times \Bigl( {\bf a}^2_2-3(a_2^{||})^2 - 2q({\bf a}_1 \cdot {\bf a}_2-3a_1^{||}a_2^{||}) \notag \\
+ q^2 ( {\bf a}_1^2 - 3(a_1^{||})^2) \Bigr) \Bigr] \hat{\bf n}_{||} - \frac{1}{9\sqrt{2}(1+q)^2} \notag \\
\Bigl( q(1+4q) {\bf a}_1 + (4+q) {\bf a}_2 \Bigr) + \frac{1}{\eta} \frac{{\bf a}_2+q^2{\bf a}_1}{(1+q)^2}. \notag
\end{align}
The other variables in these formulas are empirical fitting constants, with values found to be $E_2=.341, E_3=0.522, E_{\rm S}=0.673, E_{\Delta}=-0.3689, E_{\rm A}=-0.0136, E_{\rm B}=0.045, E_{\rm C}=0, E_{\rm D}=0.2611, E_{\rm E}=0.0959, E_{\rm F}=0, J_2=-2.81, J_3=1.69, J_{\rm A}=-2.9667, J_{\rm B}=-1.7296, J_{\Delta}=J_{\rm M}=J_{\rm S}=J_{\rm MS}=0$.  We also set $\Theta_2=\Theta_3=\Theta_4=\Theta_5=0$ because of the weak dependence of the results on these parameters.  With extremely low probability ($\sim 10^{-5}$), Eq. (B3) can give superextremal spin values, of $a'>1$; we discard these cases from our Monte Carlo sample when they appear. 

\begin{table}
\centering
\begin{tabular}{ r || r | r | r | r | r | r | r | r}
  Ref. & $q$ & $\psi$ & $a_{\rm BH}$ & $R_{\rm NS}$ & $a_{\rm BH, NR}'$ & $a_{\rm BH, PN}'$ & $\frac{M_{\rm dis}^{\rm NR}}{M_{\rm NS}}$ & $\frac{M_{\rm dis}^{\rm F}}{M_{\rm NS}}$ \\
  \hline                        
  \citep{FDK} & 1/7 & $0^{\circ}$ & 0.5 & 14.4~km & 0.67 & 0.658 & $\le 0.4\%$ & $0\%$ \\
  \citep{FDK} & 1/7 & $0^{\circ}$ & 0.7 & 14.4~km & 0.80 & 0.786 & $6\%$ & $7.2\%$ \\
  \citep{FDK} & 1/7 & $0^{\circ}$ & 0.9 & 14.4~km & 0.92 & 0.913 & $28\%$ & $22.9\%$ \\
  \citep{FDK} & 1/5 & $0^{\circ}$ & 0.5 & 14.4~km & 0.71 & 0.681 & $6\%$ & $6.5\%$ \\
  \citep{FDD12} & 1/7 & $0^{\circ}$ & 0.9 & 12.2~km & 0.923 & 0.913 & $10\%$ & $11.8\%$ \\
  \citep{FDD12} & 1/7 & $0^{\circ}$ & 0.9 & 13.3~km & 0.919 & 0.913 & $20\%$ & $17.4\%$ \\
  \citep{FDD12} & 1/7 & $0^{\circ}$ & 0.9 & 14.4~km & 0.910 & 0.913 & $30\%$ & $22.1\%$ \\
  \citep{FDD12} & 1/7 & $20^{\circ}$ & 0.9 & 14.4~km & 0.909 & 0.911 & $28\%$ & $20.3\%$ \\
  \citep{FDD12} & 1/7 & $40^{\circ}$ & 0.9 & 14.4~km & 0.898 & 0.900 & $15\%$ & $13.8\%$ \\
  \citep{FDD12} & 1/7 & $60^{\circ}$ & 0.9 & 14.4~km & 0.862 & 0.870 & $3\%$ & $1.3\%$ \\
  \citep{FDKT} & 1/3 & $0^{\circ}$ & 0.0 & 14.6~km & 0.56 & 0.54 & $5.2\%$ & $5.21\%$\\
  \citep{FDKT} & 1/3 & $0^{\circ}$ & 0.5 & 14.6~km & 0.77 & 0.70 & $15.5\%$ & $16.3\%$ \\
  \citep{FDKT} & 1/3 & $0^{\circ}$ & 0.9 & 14.6~km & 0.93 & 0.829 & $38.9\%$ & $28.3\%$ \\
  \citep{FDKT} & 1/3 & $20^{\circ}$ & 0.5 & 14.6~km & 0.76 & 0.699 & $14.5\%$ & $15.8\%$ \\
  \citep{FDKT} & 1/3 & $40^{\circ}$ & 0.5 & 14.6~km & 0.74 & 0.691 & $11.5\%$ & $14.1\%$ \\
  \citep{FDKT} & 1/3 & $60^{\circ}$ & 0.5 & 14.6~km & 0.71 & 0.671 & $8.0\%$ & $11.4\%$  \\
  \citep{FDKT} & 1/3 & $80^{\circ}$ & 0.5 & 14.6~km & 0.66 & 0.636 & $6.1\%$ & $8.0\%$ \\
          \end{tabular}
\label{PNValidation}
\caption{Comparison of the full NR results for postmerger BH spin $a_{\rm BH}'$ to the predictions of our PN fitting formula, Eq. \eqref{aPrime}.  Here $\psi$ is the initial spin-orbit misalignment angle, $q$ is the mass ratio, and the ``Ref.'' column refers to the paper whose NR data we are comparing the PN results to.  The agreement is quite strong for small $q$.  For $q=1/3$, roughly the upper limit considered in our BH mass functions, the error is $\lesssim 10\%$, which is still acceptable for our purposes.  In the last two columns we also demonstrate the reasonable agreement between NR estimates for initial postmerger disk mass, $M_{\rm d}^{\rm NR}$, and the prescription of Eq. \eqref{MdFF}, $M_{\rm dis}^{\rm F}$.  The disk mass fitting formula works very well for low and moderate values of spin but becomes less accurate for the $a_{\rm BH}=0.9$ runs.  In each simulation in this table, the NS mass is $1.4M_{\odot}$.}
\end{table}

As mentioned in Sec. III, these formulas, which were derived, and calibrated, for BH-BH mergers, are found to give surprisingly good agreement with detailed results for BH-NS mergers simulated in full NR.  We demonstrate the agreement in Table II, where we also plot the generally good agreement between Eq. \eqref{MdFF} and NR results for the initial mass of the postmerger accretion disk.  The largest errors in the disk mass fitting formula seem to occur for large disk masses ($M_{\rm dis}>0.2M_{\rm NS}$), and for high premerger spin-orbit misalignments ($\psi \gtrsim 60^{\circ}$).  The first of these cases occurs only for a small subset of our BH-NS mergers; the second occurs for a larger fraction, and may result in a modest overestimate of $f_{\rm GRB}$.

\end{document}